\documentclass[sigconf]{acmart}
\usepackage{multirow}
\usepackage{amssymb}
\usepackage[normalem]{ulem}
\usepackage{booktabs} 
\usepackage{todonotes}
\usepackage{array}
\usepackage{soul}
\newcolumntype{P}[1]{>{\centering\arraybackslash}p{#1}}

\AtBeginDocument{%
  \providecommand\BibTeX{{%
    \normalfont B\kern-0.5em{\scshape i\kern-0.25em b}\kern-0.8em\TeX}}}

\setcopyright{acmlicensed}

\copyrightyear{2019} 
\acmYear{2019} 
\acmConference[RecSys '19]{Thirteenth ACM Conference on Recommender Systems}{September 16--20, 2019}{Copenhagen, Denmark}
\acmBooktitle{Thirteenth ACM Conference on Recommender Systems (RecSys '19), September 16--20, 2019, Copenhagen, Denmark}
\acmPrice{15.00}
\acmDOI{10.1145/3298689.3347020}
\acmISBN{978-1-4503-6243-6/19/09}

\begin{document}

\title[Towards A Recommender System for Improving Wellbeing]{Aligning Daily Activities with Personality:
Towards A Recommender System for Improving Wellbeing}


\author{Mohammed Khwaja$^{1,2}$, Miquel Ferrer$^{1}$, Jesus Omana Iglesias$^{1}$, A. Aldo Faisal$^{2}$, Aleksandar Matic$^{1}$}
\affiliation{%
\vspace{0.1cm}
 \institution{$^1$Telefonica Alpha, Spain \hspace{0.3em} $^2$Brain \& Behaviour Lab, Department of Bioengineering, Imperial College London, UK}
}
\affiliation{%
\vspace{0.1cm}
  \institution{\{mohammed.khwaja, miquel.ferrer, jesus.omana, aleksandar.matic\}@telefonica.com, a.faisal@imperial.ac.uk}}

\renewcommand{\shortauthors}{Khwaja, et al.}

\begin{abstract}
Recommender Systems have not been explored to a great extent for improving health and subjective wellbeing. Recent advances in mobile technologies and user modelling present the opportunity for delivering such systems, however the key issue is understanding the drivers of subjective wellbeing at an individual level. In this paper we propose a novel approach for deriving personalized activity recommendations to improve subjective wellbeing by maximizing the congruence between activities and personality traits. To evaluate the model, we leveraged a rich dataset collected in a smartphone study, which contains three weeks of daily activity probes, the Big-Five personality questionnaire and subjective wellbeing surveys. We show that the model correctly infers a range of activities that are 'good' or 'bad' (i.e. that are positively or negatively related to subjective wellbeing) for a given user and that the derived recommendations greatly match outcomes in the real-world.
\end{abstract}



\maketitle

\section{Introduction and related work}

The pursuit of happiness is the ultimate goal for many people - described by Aristotle as the meaning and purpose of life. Interestingly, the scientific pursuit of happiness and life satisfaction (together referred to as "subjective wellbeing", or SWB) has intensified in the last quarter of the 20th century. Since then, the number of articles in the domain of SWB have grown exponentially~\cite{myers2000hope}. Now we have a wealth of knowledge~\cite{diener2018advances} on how to measure SWB and a deeper understanding on how it correlates with cognitive, behavioral, environmental, biological and genetical factors. On the other hand, interdisciplinary research has successfully demonstrated the potential of mobile computing to quantify and monitor human behaviors accurately and at a scale larger than ever before~\cite{matic2012multi, miluzzo2007cenceme, atallah2009use, srivastava2012human}, gain insight into mental wellbeing~\cite{servia2017mobile, lane2011bewell, lathia2013smartphones} and automatically predict user characteristics such as personality traits~\cite{khwaja2019ubicomp, khwaja2019interact}. The advances of mobile technologies coupled with knowledge from SWB research can open the door to proactive recommendations to help people make decisions or adapt their daily activities to maximize happiness and life satisfaction. While the field of recommender systems (RSs) has provided numerous tools to support user decision making by identifying personalized and relevant content, services or products \cite{ricci2015recommender}, RSs that provide personalized suggestions to boost SWB have not attracted a considerable research interest yet. 

One of the key issues of utilizing advances in RSs for providing SWB recommendations is the existence of appropriate datasets. Contrary to the availability of rich datasets on movies, music, books, products, etc. matched with user profiles, it is not trivial to collect equally large datasets that match precise information about users' activities and ground-truth information about their SWB. This represents a common 'cold-start' problem -- a situation of having sparse historical data or not having enough information about new users \cite{hu2011enhancing, tkalcic2015personality}. In order to tackle the cold-start problem in developing RSs for SWB, we use personality traits in this study as a proxy to user profiles, relevant for matching daily activities to subjective wellbeing at an individual level. Our approach is inspired by the congruence between personality and daily activities identified in psychology \cite{diener1999subjective} as well as by recent trends of leveraging personality traits for addressing the cold-start problem in RSs \cite{hu2011enhancing, tkalcic2009personality, elahi2013personality}. Personality has been used successfully to recommend movies \cite{karumur2016exploring, fernandez2016alleviating, wu2013using}, music \cite{hu2011enhancing, fernandez2016alleviating, ferwerda2016personality}, books~\cite{fernandez2016alleviating}, and also leisure activities and events \cite{nurbakova2017users}. However the aim of these studies was to optimize item or event selection rates rather than to optimize users' SWB. Behavioral Economics suggests that people have biases in understanding the link between behaviours and their SWB~\cite{kahneman2006developments}. Importantly, literature also suggests that it is possible to improve an individual's SWB\footnote{Despite the fact that genetical predisposition explains 50\% of SWB variance}\cite{lyubomirsky2005pursuing, diener1999subjective}. Our work is novel in providing a technological foundation for an RS to support people in improving SWB, with the following distinct contributions:
\begin{itemize}
\item A machine learning algorithm that predicts users' SWB based on the congruence between their reported Big-Five personality traits and distribution of their activities (Section~\ref{congruence}).
\item Evaluation of the predictive power of the algorithm in simulating distribution of activities that would result in low or high SWB for a specific user - thus constituting a personalized whitelist and blacklist for increasing SWB (Section~\ref{range}).
\end{itemize}

\section{In-the-wild Data Collection}
\label{dataset}


To build the dataset required for the SWB RS, we used a smartphone app that captures patterns of daily activities that people were engaged in over a period of 2-3 weeks. The onboarding survey (delivered through the smartphone app) prompted participants to answer the Big-Five personality questionnaire~\cite{goldberg2006international} and a SWB question. We used the question \textit{'Overall, how satisfied are you with your life nowadays'} as the ground-truth information for SWB, referred to as "life satisfaction"~\cite{dolan2011measuring}. During the following 2-3 weeks, we applied the Ecological Momentary Assessment method~\cite{shiffman2008ecological} by prompting participants at five random times distributed over a day to report the activity they were engaged in at that moment.  

The data collection was conducted in two trials. The first trial (\textbf{first dataset}) ran between February and August 2018 with 151 participants completing the study. Participants in this sample were involved through a recruitment agency from five different countries (UK, Spain, Colombia, Peru and Chile). The second trial (\textbf{second dataset}) was conducted between January and March 2019 with 256 participants completing the study from a major UK university. Participants in this sample were recruited using an email sent to students and staff at the university. For their effort to complete the study, participants in both samples were rewarded with a monetary contribution. 
We used the first dataset to build the model described in Section~\ref{congruence} and the second to validate the RS described in Section~\ref{range}. We considered this to be a rigorous way of testing the generality of our model than, for instance, performing cross-validation or using dataset splitting methodologies. This is attributed to the fact that the two datasets included different demographics (in terms of level of education, socio-economic status, age and nationality) and were collected during different periods of the year to account for potential seasonal effects.

\section{Modelling Congruence and its relationship with wellbeing}
\label{congruence}

\begin{figure}[]
  \includegraphics[width=0.9\linewidth]{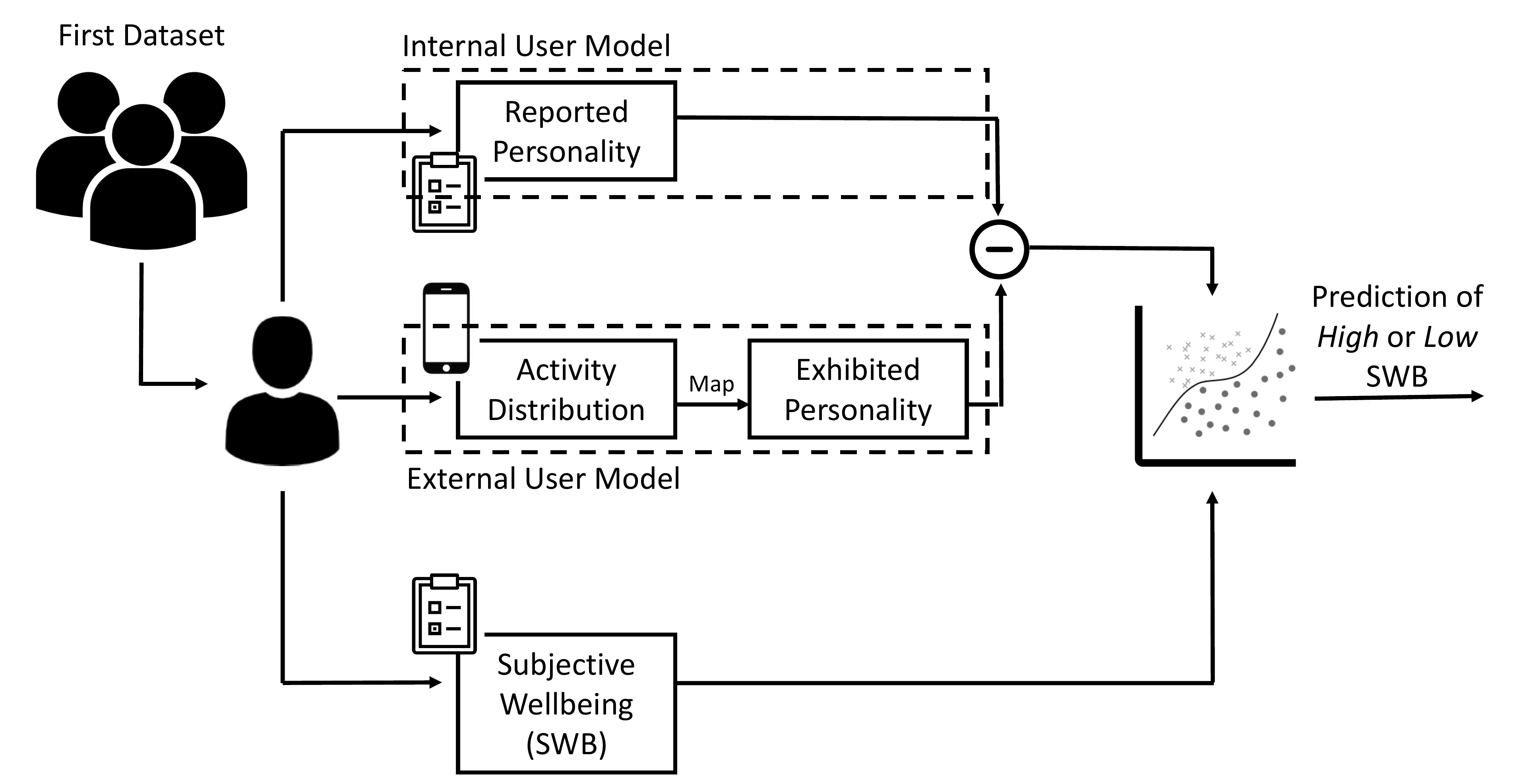}
  \caption{Classification of high/low SWB based on the congruence user model.} 
  \label{fig:model1}
  \vspace{-1.4mm}
\end{figure}

Psychologists have emphasized that the congruence between internal and external factors is a predictor of SWB~\cite{diener1999subjective}. In practice, internal factors are usually static or slow-changing (inherent to an individual's innate characteristics), whereas external factors are mostly dynamic (exhibited by an individual's behaviour and environmental circumstances). Inspired by this, we defined the \textbf{congruence user model} that quantifies the alignment between the internal and external factors (i.e. internal and external user models). Figure~\ref{fig:model1} summarizes the development process of the congruence model and it takes into account the difference between the internal and external user models to predict SWB. As an example, the external model may indicate that a user behaves as an introverted person (based on her activities) although she is an extroverted person (based on the Big-Five questionnaire). The congruence model quantifies the gap between the two. Subsequently, we explore if the machine learning model that considers this misalignment between who she is (internal model) and how she behaves (external model) predicts a lower SWB. The performance of this model is evaluated using the first dataset, and then the generated recommendations at an individual level are validated with the second dataset.


\subsection{Congruence User Model}

We used the Big-Five personality traits~\cite{goldberg2006international} to represent the user's internal model, as they are one of the most representative personal characteristics that describe and predict human behaviour~\cite{corr2009cambridge}. As personality is obtained from a self-reported questionnaire, we refer to this as the \textbf{reported personality}. Formally, for user $j$, the reported personality vector is given by: 

\begin{equation}
\vec{p}_{r}^{(j)} = <e_r^{(j)},a_r^{(j)},c_r^{(j)},n_r^{(j)},o_r^{(j)}>
\label{a}
\end{equation}
\noindent
where, $e_r^{(j)},a_r^{(j)},c_r^{(j)},n_r^{(j)},o_r^{(j)}$ represent the 5 reported personality (extraversion, agreeableness, conscientiousness, neuroticism and openness) scores respectively, for the $j^{th}$ user. 

We built a user's external user model from his/her patterns of daily activities, e.g., eating, working, watching TV, shopping, listening to music, using social media, exercising, and so on. 
Due to the variety of momentary activity items, we grouped the reported activity items into $n$ activity categories as defined by Goldberg~\cite{goldberg200911}. Goldberg conducted a 10 year long study with 800 individuals, and clustered 400 activity items into 33 categories. For simplicity, we refer to the activity categories as activities throughout the paper. To model the alignment between personality and activities, we also used Goldberg's study as the state-of-the-art dataset to provide correlations between activities and personality traits. 

First, we captured the distribution of a user's activities, which corresponds to the normalized frequencies of all the activities that the $j^{th}$ user reports. Mathematically, this is represented by a vector as: $\overrightarrow{act}^{(j)} = {<act_1^{(j)}, act_2^{(j)}, ..., act_n^{(j)}>}/{\sum_{i=1}^{n} act_{i}^{(j)}}$. Here, $act_{i}^{(j)}$ is the frequency of activity $i$ for user $j$. Moreover, the sum of all components of $\overrightarrow{act}^{(j)}$ for any user $j$ equals 1. Our external model builds a secondary personality that is \textit{exhibited} through the user's activity distribution, to directly compare against the internal model. 
We computed a dynamic construct called \textbf{exhibited personality}, $\vec{p}_{ex}^{(j)}$ that contains five dimensions similar to $\vec{p}_{r}^{(j)}$ and is modelled based on the user's activity distribution $\overrightarrow{act}^{(j)}$. For a user $j$, the exhibited personality vector is given by: 

\begin{equation}
\vec{p}_{ex}^{(j)} = f(\overrightarrow{act}^{(j)}) = <e_{map}^{(j)},a_{map}^{(j)},c_{map}^{(j)},n_{map}^{(j)},o_{map}^{(j)}> 
\label{b}
\end{equation}
\noindent

\noindent where, $e_{map}^{(j)},a_{map}^{(j)},c_{map}^{(j)},n_{map}^{(j)},o_{map}^{(j)}$ represent the 5 mapped personality scores, as a function of $f$. To obtain $f$, we first acquired a weight vector $\vec{w}^{(j)}$ that defines the positive/negative accumulated effect (or weight) of each activity on the traits. By using the correlation matrix between activities and personality defined in Goldberg~\cite{goldberg200911}, represented as $\mathbf{C}$, we define the weight of \textit{exhibited} activities on the personality as: $\vec{w}_{(j)} = \mathbf{C} \cdot \overrightarrow{act}^{(j)}$. Here, $\cdot$ is the matrix product and  $\vec{w}_{(j)}$ is obtained by using $\overrightarrow{act}^{(j)}$ as a column matrix. Using this, we derived the weighted median personality that represents the change over or below the median personality exhibited by one's activity patterns. Thus, $\vec{p}_{ex}^{(j)}$ is obtained as the vector sum of the median personality $\vec{p}_{median}$ and the weighted median personality $\vec{p}_{median} \odot \vec{w}_{(j)}$:

\begin{equation}
\begin{split}
\vec{p}_{ex}^{(j)} & = \vec{p}_{median} + \vec{p}_{median} \odot \vec{w}_{(j)}\\
& = \vec{p}_{median} \odot (1 + \mathbf{C} \cdot \overrightarrow{act}^{(j)})
\end{split}
\label{d}
\end{equation}
\noindent

\noindent indicating that $f(\vec{x}) =  \vec{p}_{median} \odot (1 + \mathbf{C} \cdot \vec{x})$, where $\odot$ is the Hadamard product. Finally, the congruence user model of a person is the difference between $\vec{p}_{r}^{(j)}$ and $\vec{p}_{ex}^{(j)}$, normalised by $\vec{p}_{r}^{(j)}$. This difference, or \textit{delta}, is given by:

\begin{equation}
\vec{p}_{\Delta}^{(j)} = {(\vec{p}_{r}^{(j)} - \vec{p}_{ex}^{(j)})}/{\vec{p}_{r}^{(j)}} = <e_{\Delta}^{(j)}, a_{\Delta}^{(j)}, c_{\Delta}^{(j)}, n_{\Delta}^{(j)}, o_{\Delta}^{(j)}>
\label{e}
\end{equation}
\noindent where $e_{\Delta}^{(j)}, a_{\Delta}^{(j)}, c_{\Delta}^{(j)}, n_{\Delta}^{(j)}, o_{\Delta}^{(j)}$ are the delta components of extraversion, agreeableness, conscientiousness, neuroticism and openness respectively. As each component of $\vec{p}_{\Delta}^{(j)}$ decreases, more congruent is the user's behaviour with respect to his/her personality; and as per our hypothesis, more is the SWB. 

\begin{table}
 \caption{Classification of SWB}
     \vspace{-1.5mm}
    \centering
    \begin{tabular}{P{4.1cm}|P{1.2cm}|P{0.9cm}|P{0.8cm}}
    \hline
    Model Type (Features) & Accuracy & Kappa & AUC\\\hline
    Personality ($\vec{p}_{r}^{(j)}$) & 57\% & 0.11 & 0.55\\
    Activity ($\overrightarrow{act}^{(j)}$) & 52\% & 0.04 & 0.52\\
    Personality-Activity ($\vec{p}_{r}^{(j)}, \overrightarrow{act}^{(j)}$) & 61\% & 0.18 & 0.58\\
    Congruence ($\vec{p}_{\Delta}^{(j)}$) & \textbf{70}\% & \textbf{0.38} & \textbf{0.71}\\
    \hline
    \end{tabular}
    \label{tab:mlmodel}
    \vspace{-3.5mm}
\end{table}

\subsection{Experimentation and Results}
\label{classification}

We built a machine learning model that predicts the SWB score for a user $j$, by using the individual \textit{delta} scores along the five personality dimensions, $\vec{p}_{\Delta}^{(j)}$. For preforming this analysis, we use the first dataset described in Section~\ref{dataset}. We cluster momentary activity items into 15 activity categories from the 33 defined in~\cite{goldberg200911}, that are most relevant to items reported in the dataset. The values of each component of $\vec{p}_{r}$ vary from 10-50 and SWB is rated on a scale from 1-10. We treat the prediction of SWB as a binary classification problem by dividing the continuous variable into high (1) and low (0), using the median value as the threshold. We use the leave-one-sample-out method to evaluate the the model accuracy. We tested different machine learning algorithms, namely: random forest, na\"ive bayes and support vector machine (SVM), and observed that the latter performs the best. For brevity, we report results with SVM only. To assess the added value of the congruence user model, we compared it to classifiers that used only 
personality traits $\vec{p}_{r}^{(j)}$, only activity distributions $\overrightarrow{act}^{(j)}$ and the combination of the two $\{\vec{p}_{r}^{(j)}, \overrightarrow{act}^{(j)}\}$, shown in Table~\ref{tab:mlmodel}. We observed that the classifier relying on the congruence user model and the computed \textit{delta} features outperforms the other classifiers in predicting SWB. 

This illustrates the strength of incorporating the congruence theory into the SWB classification model, in comparison to using a typical "black-box" model that relies on the same inputs -- personality and activity distributions. By relying on this model, an RS to improve a user's SWB would aim to suggest activity distributions that improves the personality-activities alignment, i.e. reduce the gap quantified through \textit{delta} features. 
Certainly, there is more than one distribution of activities that corresponds to low values of \textit{delta} features for a given user, and the importance of minimizing all of the five \textit{delta} values (i.e. alignment with all the five personality traits) is not equally important. In our experiments, we observed that the relative ratio among different activities and also among \textit{delta} features matters more than the absolute values of activity frequency or the \textit{delta} values. This also corroborates with an intuitive assumption that there is no one unique lifestyle beneficial for an individual - i.e. understanding the range of activity distributions is beneficial to improving a user's SWB. 


\section{Activity Range Recommendation}
\label{range}
\begin{figure}
  \includegraphics[width=0.9\linewidth]{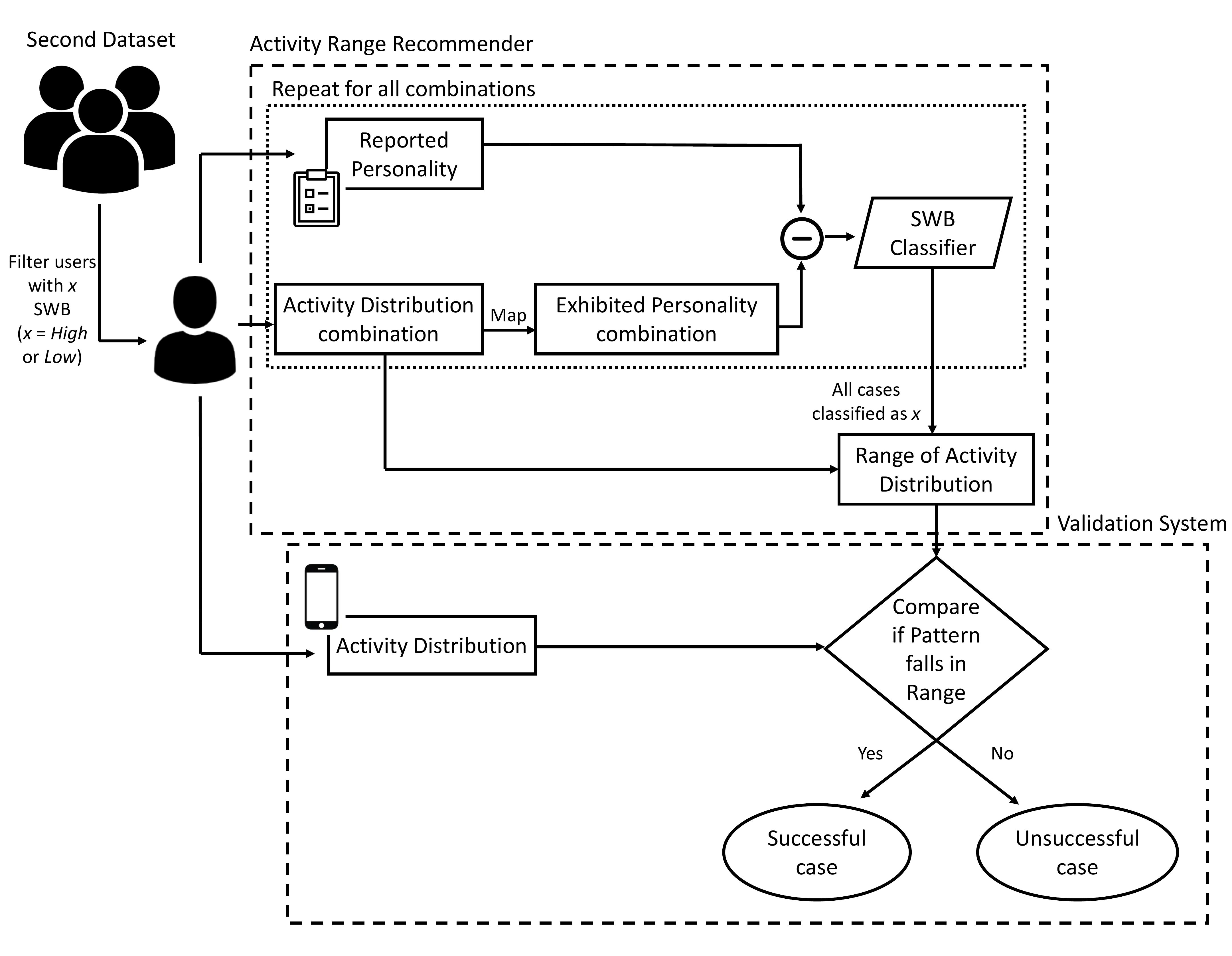}
  \caption{Model for the activity range recommendation, along with the validation procedure}
  \label{fig:model2}
\end{figure}

In this section we describe a methodology to recommend a range of relevant activities that are 'good' or 'bad for a user, and subsequently test the results using the second dataset to assess if the model scales to a different population. The flow diagram of the RS, along with the validation procedure is summarized in Figure~\ref{fig:model2}. Firstly, the activity range recommender compares different exhibited personalities for all simulated combinations of activity distributions with a user's reported personality that will result in high or low SWB, using the SWB classifier described in Section~\ref{classification}. Through this process we obtain the range of activity distributions that form the whitelist (good) and the blacklist (bad) for the user's SWB. The whitelist recommendation is validated by comparing against the actual activity distribution for users in the second dataset that have high SWB. The same is done for the blacklist recomendation with users that have low SWB. 

\subsection{Methodology}
Though the sum of all components of $\overrightarrow{act}^{(j)}$ equals 1 for a user $j$, i.e., $\sum_{i=1}^{n} \langle\overrightarrow{act}_{i}^{(j)}\rangle = 1$, some of these components vary significantly across the sample, while others show a lower variance
. The activities that do not vary across the sample also do not predominantly impact the calculated values of exhibited personality $\vec{p}_{ex}^{(j)}$. It is also intuitively clear that there are certain activities (usually those that we have less control of) that occupy a similar proportion of time spent by most people (such as working, studying, sleeping, eating, etc.). Hence, to narrow down the list of recommended activities (such as using social media, watching TV, reading, exercising, etc.) that may be more actionable (i.e. more under a user's control) for providing recommendations, we consider only those activities that have high variance in the sample. 
Without loss of generality, it is assumed that there are $m$ activities that have high variance from $n$ activities (with $m \leq n$) so that $\sum_{i=1}^{m} \langle\overrightarrow{act}_{i}^{(j)}\rangle = 1-\lambda$. Here, $\lambda$ is a relatively low value ($\sim$0.2) that covers the joined variance of the $n-m$ activities. For the selected $m$ activities, we obtained all potential combinations in increments of 0.1 such that this condition is met. For each of these combinations, we calculate $\vec{p}_{ex}^{(j)}$ and compute the \textit{delta}, $\vec{p}_{\Delta}^{(j)}$ for each user $j$ by comparing against $\vec{p}_{r}^{(j)}$. We predict the SWB for all $\vec{p}_{\Delta}^{(j)}$ possibilities using the SWB classifier described in Section~\ref{congruence}, and marked the cases where this is high. Using these, we determined the range (sorted from lowest to highest) of distributions for each of the $m$ activities that are expected to give high SWB. We performed the same procedure for the cases of low SWB, indicating the range of activity distributions that result in low SWB.

\subsection{Validation and Results}

We use the second dataset described in Section~\ref{dataset} to validate the hypothetical outputs of our RS. As with the first dataset, we use the same $n=15$ categories to cluster the activities. We observed that $m=8$ clusters have significant variance ($>0.1$) across the sample and have the most effect on $\vec{p}_{ex}^{(j)}$, and in this dataset, the value of $\lambda = 0.1$. Using the SWB report of users in the second dataset, we divide each user into either high or low class - if they are either over or under the median value respectively (as done previously with the first dataset). For each of the users in the high SWB class and low SWB class, we obtained the range of activity distributions that would provide high and low SWB respectively, according to our model. We evaluated the extent to which the activity proportions of the user fall within the range of all selected activities (all 8 activities) or majority (at least 5) of them. These results are reported in Table~\ref{tab:validation}. It is important to note that testing an activity RS proactively would be a difficult endveour as behaviour change is a complex task, and evaluating the effectiveness of interventions is particularly difficult for health and wellbeing~\cite{campbell2007designing}. Contextual factors, external events and sense of autonomy play a large role in determining the extent to which a recommendation will be followed and performed by users~\cite{deci1987support}. Hence, we evaluated the effectiveness of the RS for both an ideal case (user performs all activities as per the recommendation) and a realistic case (user performs the majority of the activities).

When considering all the activities that are important for high \& low SWB, 51\% of users in the high SWB class fall in the range, while for low SWB class the number increases to 74\%. This indicates that our method is able to infer the range of activity distributions that are beneficial for user's SWB fairly well, and furthermore it is more successful in inferring the ranges of activities that have negative consequences for SWB. When including the majority of the activities (5), these numbers improve significantly - rising to 71\% for users in the high SWB class and 92\% in the low SWB class. This resonates well with real-life scenarios as users may not be able to maintain the exact proportion for all activities all the~time.

\begin{table}
 \caption{Validation of Optimal Activity Range Recommendation System}
 \vspace{-1.5mm}
    \centering
    \begin{tabular}{P{2.3cm}|P{2.3cm}|P{2.3cm}}
    \hline
    Activities Considered & Sample Under Consideration & Predicted - Actual Accuracy\\\hline
    \multirow{2}{*}{All (8)} & High SWB & 51\%\\
     & Low SWB & 74\%\\
    \hline
     \multirow{2}{*}{Majority ($\geq5$)} & High SWB & 71\%\\
     & Low SWB & 92\%\\
    \hline
    \end{tabular}
    \label{tab:validation}
    \vspace{-5mm}
\end{table}
\vspace{-1.5mm}
\section{Conclusion}

In this paper, we presented a novel approach to developing an RS that supports users in balancing daily activities to improve their subjective wellbeing (SWB). Developing an SWB RS is a challenging task, as it faces the cold-start problem due to the lack of large-scale datasets that map user characteristics, daily activities and ground-truth information about SWB. We addressed this limitation by collecting a dataset with the above information and developed a user model based on psychological literature suggesting that the congruence between internal and external factors impacts SWB \cite{diener1999subjective}. Using this model, we built a binary classifier that predicts SWB based on the alignment between an individual's behavior (in terms of distribution of activities) and his/her personality. Our model outperformed the three benchmarking classifiers that relied on the same input parameters, by 9-18\%. Subsequently, we simulated the range of activity distributions that would result in high or low SWB for users in another dataset, and compared personalized recommendations (i.e. white- and black- listed activity distributions) to the ground-truth of users' activity distributions. Our model inferred the range of activities that are 'good' or 'bad' for a given user with accuracy up to 92\%, demonstrating that the model successfully captures the link between SWB and the alignment between real-life activity patterns with personality. We believe that this work will encourage more research, both in RSs devoted to SWB, and in psychology, aiming to deepen understanding of the congruence between individual's daily activities and personality. 

In addition to its development, testing the impact of an RS of this nature is a challenge, as the inference and provision of perfectly aligned activity distributions to one's SWB does not guarantee that a user will follow the recommendations. For this reason, we have created a tool that suggests daily activities to promote SWB. We plan to test this tool in the near-future and explore the change in SWB achieved through the suggestions provided by our system.

\section*{Acknowledgements}
This work has been supported from funding awarded by the European Union's Horizon 2020 research and innovation programme, under the Marie Sklodowska-Curie grant agreement no. 722561.

\bibliographystyle{ACM-Reference-Format}

\end{document}